\documentclass[twocolumn,showpacs,floats,floatfix,superscriptaddress,aps,pra]{revtex4-1}

\usepackage{amsfonts}
\usepackage{amssymb}
\usepackage{amsmath}
\usepackage{calc}
\usepackage{dcolumn}
\usepackage{graphicx}
\usepackage{bm}
\usepackage{epstopdf}
\usepackage{epsfig}
\usepackage{float}
\usepackage {color}
\usepackage{textcomp}
\usepackage{soul}
\usepackage{diagbox}

\DeclareMathOperator{\Tr}{Tr}

\begin{document}

\title{Broadband polarization rotator with tunable rotation angle composed of three wave-plates}
\date{\today }

\author{Mouhamad Al-Mahmoud}

\affiliation{Department of Physics, Sofia University, James Bourchier 5 blvd, 1164 Sofia, Bulgaria}

\author{Virginie Coda}
\affiliation{Universit\'e de Lorraine, CentraleSup\'elec, LMOPS, F-57000 Metz, France}

\author{Andon Rangelov}
\affiliation{Department of Physics, Sofia University, James Bourchier 5 blvd, 1164 Sofia, Bulgaria}

\author{Germano Montemezzani}
\affiliation{Universit\'e de Lorraine, CentraleSup\'elec, LMOPS, F-57000 Metz, France}

\begin{abstract}
A simple scheme for a broadband polarization rotator with tunable rotation angle is proposed and verified experimentally. The rotator consists of only three wave-plates, one of which is a full-wave plate. The robust approach inspired by the composite pulses analogy allows to compensate the wave-plate dispersion in large extent.
\end{abstract}
\maketitle



\section{Introduction}

Components able to convert the polarization state of a light wave are key elements for several optical devices and applications  \cite{Azzam,Goldstein,Duarte,Kliger-book}. In the case where broadband or tunable light sources are used, a robust and wide bandwidth operation of such components becomes of crucial importance. As an example we may consider wave-plate retarders, which are based on the birefringence properties of specific crystals and are intrinsically highly dispersive components with a strong wavelength dependence of the phase retardation. Methods to realize compact achromatic (broadband) retarders have been therefore the matter of interest for several decades \cite{West49,Destriau49,Pancharatnam55-1,Pancharatnam55-2,McIntyre68,Ardavan07,Ivanov2012,Peters2012,Dimova2014,Dimova2016}. Notably, the recent advances leading to improved broadband performance \cite{Ardavan07,Ivanov2012,Peters2012,Dimova2014,Dimova2016} take advantage of the formal analogy between the equations describing the change of polarization in the birefringent plate and the Schr\"odinger equation for the quantum-state dynamics of coupled two-level systems, pointed out first by Ardavan \cite{Ardavan07}. The related transfer of concepts allows to apply the fault-tolerant composite pulses approach \cite{Levitt86} widely used in the field of nuclear magnetic resonance (NMR) to polarization optics.
To this line of ideas belong also the recent studies of achromatic polarization rotators using a stack of several wave-plates \cite{Rangelov15,Dimova15,Stojanova2019} that we will discuss below.

An optical polarization rotator is an element that rotates the polarization of a linearly polarized input wave by a fixed angle which is independent of the input polarization direction. Probably the most commonly used polarization rotators are Faraday rotators which rely on the circular birefringence induced by a magnetic induction field (Faraday effect) \cite{Moeller-book}. Faraday rotators have the useful property of being non reciprocal with respect to reversing the propagation direction, they are therefore often used as optical isolators in connection with a polarizer and an analyzer. However, Faraday rotators are generally quite bulky and expensive and are not broadband. They suffer from the strong dispersion and temperature dependence of the Verdet constant connected to the Faraday effect. As an alternative, polarization rotators may be realized using crystals, such as quartz, exhibiting a natural optical activity (circular birefringence). For these commercially available elements the effect is reciprocal and the rotation angle is governed by the plate thickness and cannot be tuned. Also here the rather strong dispersion of the optical activity limits the useful bandwidth. Yet another approach makes use of twisted nematic liquid crystal cells \cite{Zhuang2000,Chung18} by much the same principle which is also at the base of liquid crystal display technology. Provided that the cell is sufficiently thick the polarization can follow adiabatically the local reorientation of the liquid crystal molecules. This effect is not wavelength specific and nearly achromatic components can be obtained. However, the rotation angle cannot be easily modified and thermal effects may affect the performance for high powers.

The most versatile way to realize tunable polarization rotators rely on the combination of several birefringent wave-plates (WPs). It is well known that two half-wave plates (HWPs) with their fast axes making an angle $\alpha/2$ lead to a rotator that turns the polarization by an angle $\alpha$. It is worth noticing that, if for this purpose one uses Fresnel rhombs instead of birefringent retarders, one can obtain a nearly achromatic but very bulky rotator \cite{Messaadi}. Indeed, a Fresnel rhomb does not rely on birefringence, its retardation is due to the different phase jump suffered by the $p$- and the $s$-polarized wave upon total reflection. Since these phase jumps vary only little with wavelength such a device can be broadband. Nevertheless, besides their size, another drawback of Fresnel rhombs is their strong sensitivity to the angle of incidence, which limits the angular aperture \cite{Fabricius}. Therefore for compact and tunable rotators the use of stacks of birefringent wave-plates remains the best choice. In this context Ye \cite{Ye} has considered a combination of three WPs, a variable WP retarder sandwiched between two crossed quarter-wave plates (QWPs). If the fast axis of the intermediate variable retarder bisects those of the QWPs this arrangement leads to a tunable polarization rotator for which the rotation angle $\alpha$ is half the retardation of the retarder. This concept was also used by Davis et al. \cite{Davis} to demonstrate two-dimensional electrically addressed polarization encoding by replacing the spatially homogeneous retarder by a parallel-aligned liquid-crystal spatial light modulator. However, in both cases the robustness of these devices with respect to a change of wavelength was not discussed. In principle any rotator composed by a combination of WPs should become broadband if every single WP is broadband. Along this main idea several studies of composite rotators based on the quantum-optical analogies mentioned above were performed. Rangelov and Kyoseva \cite{Rangelov15} have proposed a broadband composite polarization rotator based on the combination of two effective broadband HWPs, each of which is composed by a number of HWPs by the composite approach. The expected device performance was analyzed theoretically in terms of the so called fidelity (see below) for a total number of HWPs between 6 and 18. This concept was demonstrated experimentally in Ref. \cite{Dimova15} for 6 and 10 HWPs. Besides for the broadband configuration, also a narrowband configuration was implemented by another choice of the orientations of the individual HWP fast axes. This concept was developed further in a recent paper \cite{Stojanova2019}, where an even number of HWPs (up to 10) oriented at predetermined angles was used and the broadband behavior was tested through the transmission of a white light source through an analyzer placed after the HWP stack.

While the above approaches generally require a rather large number of WPs to achieve a sufficiently broadband operation ($\geq 6$), in the present work we consider a simplified arrangement involving only three WPs. With respect to the approach of Ye \cite{Ye}, which is also based on three WPs, here the middle retarder is replaced by a full-wave plate (FWP) and the two outer QWPs are replaced by HWPs. The concept exploits additionally the freedom of rotation of all three
elements in the row. It is shown that such a simple stack leads to a broadband polarization rotator provided that the intermediate wave-plate is placed in such a way as to counteract the dispersion of the HWPs. The rotator is robust against the initial polarization direction and the rotation angle can be tuned by rotating only one of the WPs. In Section II we give the theoretical background for the taken approach and compare it with the cases where only two HWPs would be used or the case where the intermediate full-wave plate would be placed under a "wrong" direction (phase shift of $+2\pi$ rather than $-2\pi$).  Notably, our sequence HWP (phase-shift of $+ \pi$) - FWP ($-2\pi$) - HWP ($+\pi$) is reminiscent of schemes used to address fault-tolerant coherent population transfer or phase gates in two-state quantum systems using effective zero-area pulses or combination of pulses with a vanishing total area \cite{Vasilev06,Torosov14}.
An important feature consists in the fact that,  unlike for the case of a two HWPs rotator, the three WPs approach leads to a near unity degree of linear polarization of the output wave even at wavelengths very far from the central design wavelength. In Section III we verify successfully the concept experimentally either by using a monochromatic wave at the central design wavelength or a broadband source covering a bandwidth of approximately 400 nm. Theoretical expectation and experimental results agree very well and confirm that such a simple arrangement is suitable for broadband operation.


\section{Theory}

\begin{figure}
  \includegraphics[width=\columnwidth]{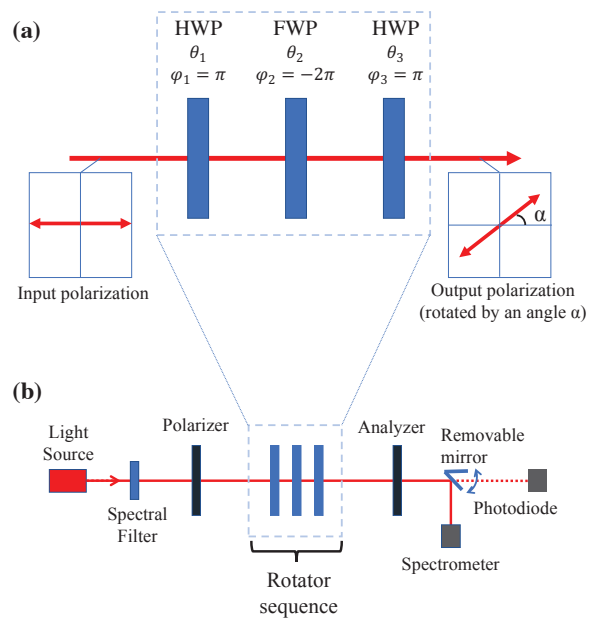}
  \caption{(a) Principle of the proposed composite polarization rotator composed of three wave-plates, a half-wave plate (HWP) followed by a full-wave plate (FWP) and another HWP. The angles $\theta_i$ are the orientation angles of each wave-plate and $\varphi_i$ are the corresponding retardations. (b) Experimental set-up for the characterization of the composite rotator.}
  \label{fig:setup-1}
\end{figure}

The broadband polarisation rotator proposed in this work is composed of three wave-plate (WP) retarders as shown in Fig.~\ref{fig:setup-1}(a). The first and the third WPs are half-wave plates for the central target wavelength $\lambda_0$ of the device, while the intermediate wave-plate is a full-wave plate at the same wavelength. Even though this element leaves the wave unchanged and acts as a neutral element at the wavelength $\lambda_0$, the importance of this crucial element for the broadband behavior will become clear below.

As it is well known, a wave-plate retarder is a birefringent element which adds different phases $\varphi/2$ and $-\varphi/2$ to two perpendicular linear polarization components of the light propagating through it. In the framework of Jones calculus \cite{Azzam,Goldstein} and in the HV-basis formed by the Jones vectors for horizontal and vertical linear polarizations, the Jones matrix for a retarder whose slow and fast axes are aligned along the HV-axes is given as
\begin{equation}
\mathcal{J}_{0}\left( \varphi \right) =\left[
\begin{array}{cc}
e^{i\varphi /2} & 0 \\
0 & e^{-i\varphi /2}%
\end{array}%
\right] .
\label{eq: retarder hv}
\end{equation}%
Here
\begin{equation}
\varphi =2\pi L(n_{\mathnormal{s}}-n_{\mathnormal{f}})/\lambda
\label{eq:retardation}
\end{equation}%
is the wave-plate retardation, i.e. the phase shift between the two orthogonal polarization components upon passing the element. The quantities $n_{\mathnormal{f}}$ and $n_{\mathnormal{s}}$ are the refractive indices along the
fast and slow axes, respectively, $\lambda$ is the vacuum wavelength of the light and $L$ is the thickness of the retarder
plate. The most commonly used retarders are the HWPs ($\varphi
=\pm \pi$) and the QWPs ($\varphi = \pm \pi /2$). A full-wave plate (FWP) has a retardation of $\varphi = \pm 2\pi$.

If the retarder plate is turned by an angle $\theta$ around the light propagation axis (perpendicular to the plate),
then its Jones matrix $\mathcal{J}_{\theta}\left( \varphi \right)$ in the HV basis can be found from $ \mathcal{J}_{0}\left( \varphi \right)$ as
\begin{eqnarray}
\mathcal{J}_{\theta}\left( \varphi \right) &=& \mathcal{R} \left( -\theta \right) \mathcal{J}_{0}\left( \varphi \right) \mathcal{R} \left( \theta \right),
\label{eq: retarder rotated}
\end{eqnarray}%
where $\mathcal{R} \left( \theta \right)$ is an axes rotation matrix given as
\begin{equation}
\mathcal{R} \left( \theta \right) =\left[
\begin{array}{cc}
\cos \theta  & \sin \theta  \\
-\sin \theta  & \cos \theta
\end{array}%
\right] .
\label{eq: rotation-matrix}
\end{equation}%
Explicitly the form of $\mathcal{J}_{\theta}\left( \varphi \right)$ is
\begin{eqnarray}
\mathcal{J}_{\theta}\left( \varphi \right)_{11} &=& e^{i\varphi /2}\cos^2{\left( \theta \right)}+e^{-i\varphi /2}\sin^2{\left( \theta \right)},\\
\mathcal{J}_{\theta}\left( \varphi \right)_{12} &=& i\sin{\left( 2\theta \right)}\sin{\left( \varphi/2 \right)},\\
\mathcal{J}_{\theta}\left( \varphi \right)_{21} &=& i\sin{\left( 2\theta \right)}\sin{\left( \varphi/2 \right)},\\
\mathcal{J}_{\theta}\left( \varphi \right)_{22} &=& e^{-i\varphi /2}\cos^2{\left( \theta \right)}+e^{i\varphi /2}\sin^2{\left( \theta \right)} .
\end{eqnarray}%

It is worth noticing that (up to an unimportant minus sign) the rotation matrix (\ref{eq: rotation-matrix}) corresponds to the Jones matrix $\mathcal{J}_\mathcal{R}\left(\alpha \right)$ of an optical rotator in the HV basis for a polarization rotation by $+\alpha$ (with the positive angles defined in counterclockwise direction), indeed
\begin{equation}
\mathcal{J}_\mathcal{R}\left(\alpha \right)=\mathcal{R} \left( -\alpha \right) =\left[
\begin{array}{cc}
\cos \alpha  & -\sin \alpha  \\
\sin \alpha  & \cos \alpha
\end{array}%
\right] .
\label{eq: rotator hv}
\end{equation}%

To simplify the above matrices $\mathcal{J}_{\theta}\left( \varphi \right)$ and $\mathcal{J}_\mathcal{R}\left(\alpha \right)$, it is convenient to express them in the left-right circular polarization basis (LR), where the basis Jones vectors are $L=1/\sqrt{2}(1, i)$ (left circular polarized) and $R=1/\sqrt{2}(1, -i)$ (right circular polarized). We call the corresponding Jones matrices in this basis $J_\theta\left( \varphi \right)$ and $J_R\left(\alpha \right)$ (note the different symbol with respect to the HV basis), a straightforward calculation leads to
\begin{equation}
J_R\left( \alpha \right) =\left[
\begin{array}{cc}
e^{-i \alpha} & 0 \\
0 & e^{i \alpha}%
\end{array}%
\right] ,
\label{eq: rotator lr}
\end{equation}%
and
\begin{equation}
J_{\theta}\left( \varphi \right) =\left[
\begin{array}{cc}
\cos \dfrac{\varphi}{2}  & \mathnormal{i e}^{-2\imath \theta}\sin \dfrac{\varphi}{2} \\
\mathnormal{i e}^{2\imath \theta}\sin \dfrac{\varphi}{2}  & \cos \dfrac{\varphi}{2}
\end{array}%
\right] .
\label{eq:Jones seq mat}
\end{equation}%

For our sequence of HWP-FWP-HWP shown in Fig.~\ref{fig:setup-1}(a) we have $\varphi_1=\pi$, $\varphi_2=\pm 2\pi$ and $\varphi_3=\pi$, the corresponding orientations for the three wave-plates are $\theta_1$, $\theta_2$ and $\theta_3$, respectively. The overall Jones matrix describing this composite sequence is therefore
\begin{equation}
J=J_{\theta_3}\left( \pi \right) J_{\theta_2}\left( \pm2\pi \right) J_{\theta_1}\left( \pi \right)  ,
 \label{eq:Jones rot seq mat-1}
\end{equation}
which gives
\begin{equation}
J=\left[
\begin{array}{cc}
\mathnormal{e}^{-2i(\theta_3 - \theta_1)}  & 0 \\
0  & \mathnormal{e}^{2i(\theta_3 - \theta_1)}
\end{array}%
\right]
=\left[
\begin{array}{cc}
e^{-i \alpha} & 0 \\
0 & e^{i \alpha}%
\end{array}%
\right] .
 \label{eq:Jones rot seq mat-2}
\end{equation}
Obviously the last equality shows that $J$ corresponds to the rotator matrix in the LR basis in Eq.~(\ref{eq: rotator lr}), therefore this sequence acts as a rotator with a rotation angle
\begin{equation}
\alpha=2(\theta_3 - \theta_1) \ .
\label{eq: alpha of thetas}
\end{equation}
This equivalence is exact at the central wavelength, for which the retardations  $\varphi_1$, $\varphi_2$ and $\varphi_3$ correspond exactly to those given above. However, we are principally interested in the behavior found when these retardations depart from the values  $\pi$, $\pm 2\pi$ and $\pi$, as a result of using a different wavelength (see Eq.~(\ref{eq:retardation})). Therefore, in order to explore the behavior in the $\varphi$ space, we define the fidelity factor according to
\begin{eqnarray}
\mathcal{F}\equiv \dfrac{1}{2} \Tr \left[J_R^{-1}\left(\alpha \right) J\right] ,
\end{eqnarray}%
where we note that $J_R^{-1}\left(\alpha \right)=J_R\left(-\alpha \right)$. The fidelity $\mathcal{F}$ is therefore a kind of measure on how close the composite matrix $J$ approaches the target matrix $J_R(\alpha)$. In the case where the output light maintains a linear polarization state, the fidelity $\mathcal{F}$ finds a more direct physical interpretation, as will be mentioned later in the experimental section.
\begin{figure}[t!]
\includegraphics[width=\columnwidth]{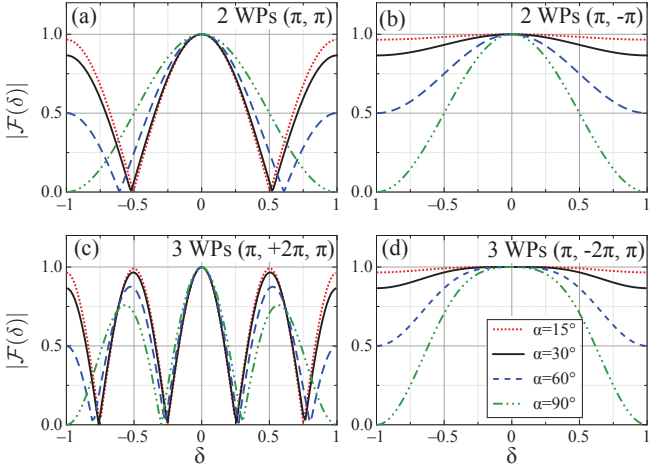}\\
\caption{Absolute value of the fidelity $|\mathcal{F}|$ as a function of the relative retardation deviation $\delta$ for four different optical rotator configurations. (a) Two HWP with their fast axes making an angle of $\alpha/2$, equivalent to $\varphi=\pi$ for both wave-plates at the central wavelength. (b) Two HWP with the fast axis of the first making an angle $\alpha/2$ with the slow axis of the second, equivalent to $\varphi_1=\pi=-\varphi_2$. (c) Three wave-plates, a FWP sandwiched between two HWPs. The fast axis of the FWP makes an angle of $+\alpha/4$ ($-\alpha/4$) with respect to the fast axis of the first (third) waveplate. The retardations are $\varphi_1=\pi$, $\varphi_2=+ 2\pi$ and $\varphi_3=\pi$. (d)  Same as (c) but the fast and slow axes of the FWP are switched (see Eq.~(\ref{eq: angles relations})), here $\varphi_1=\pi$, $\varphi_2=-2\pi$ and $\varphi_3=\pi$.  The curves are for following target rotation angles: $\alpha=15$ deg (dotted red line), $\alpha=30$ deg (solid black line), $\alpha=60$ deg (dashed blue line) and $\alpha=90$ deg (dashed-dotted green line).}
  \label{fig:fidelity}
\end{figure}

\begin{figure}[h]
\includegraphics[width=\columnwidth]{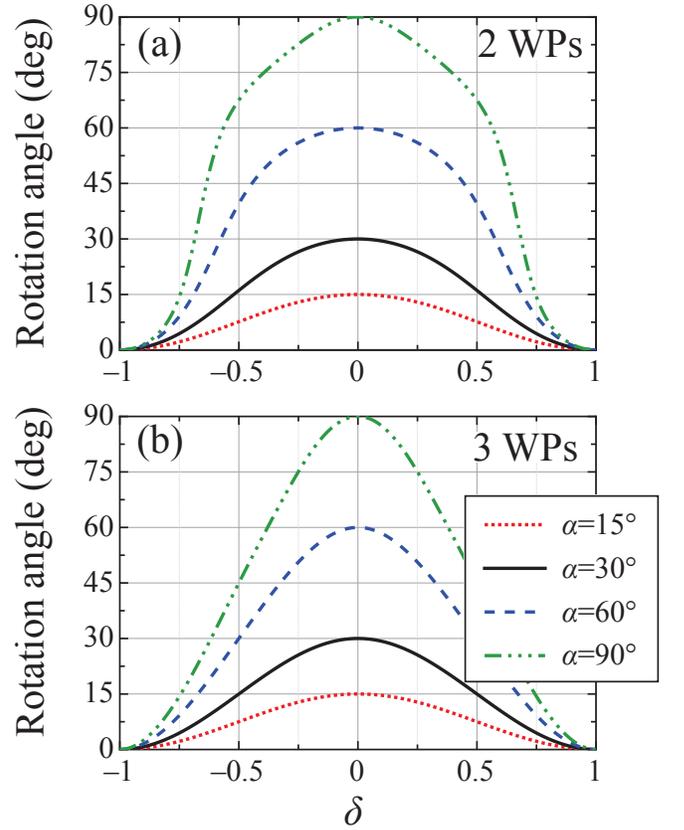}
    \caption{Output polarization rotation angle as a function of the relative retardation deviation $\delta$. (a) Sequence of two HWP ($\varphi_1=\pi$, $\varphi_2=-\pi$) as in Fig.~\ref{fig:fidelity}(b); (b) Sequence of three wave-plates ($\varphi_1=\pi$, $\varphi_2=-2\pi$ and $\varphi_3=\pi$) as in Fig.~\ref{fig:fidelity}(d). The curves for four different target rotation angles $\alpha$ have the same styles as in Fig.~\ref{fig:fidelity}. }
  \label{fig:rot-angle_theory}
\end{figure}

Let us consider the relative deviation $\delta$ from the retardation values  $\varphi_i=\varphi(\lambda_0)$ of the three wave-plates defined as
\begin{equation}
 \delta \equiv \frac{\varphi(\lambda)} {\varphi(\lambda_0)}-1= \frac{\Delta n(\lambda)}{\Delta n(\lambda_0)}\frac{\lambda_0}{\lambda}-1\ ,
\label{eq: def delta}
\end{equation}
where $\lambda_0$ is the central wavelength for which the composite structure is designed and $\Delta n=n_s-n_f$ in Eq.~(\ref{eq:retardation}).

While the orientation angles $\theta_1$ and $\theta_3$ are related by Eq.~(\ref{eq: alpha of thetas}), the optimum angle $\theta_2$ can be found by maximizing the integral of the fidelity $\mathcal{F}$ over the range $-1 \leq \delta \leq 1$,  i.e.
\begin{equation}
Q =  \frac{1}{2} \int_{-1}^{+1} |\mathcal{F}(\delta)| d\delta  \ ,
\label{figure_of_merit}
\end{equation}
which means searching for the broadest fidelity curves $\mathcal{F}(\delta)$. The above integral can be considered as a quality measure (figure-of-merit) for the robustness of the rotator and its maximum possible value equals 1.
It can be shown analytically that this integral is maximized if
\begin{equation}
\theta_2 =  \theta_1 + \alpha/4 -\pi/2 =  \theta_3 - \alpha/4 -\pi/2 \ .
\label{eq: angles relations}
\end{equation}
The additional angle $\pi/2$ appearing above is important. In fact, the analysis was performed by assuming $\varphi_2(\lambda_0)=+2\pi$, however, the additional angle of $\pi/2$ implies that the role of the slow and fast axis should be permuted for the intermediate FWP, meaning that the optimum is found for a negative retardation $\varphi_2(\lambda_0)=-2\pi$ (by simultaneously dropping the $-\pi/2$ term in Eq.~(\ref{eq: angles relations})).

To illustrate the expected robustness of the composite rotator we depict in Fig.~\ref{fig:fidelity} the expected fidelity as a function of the relative retardation deviation $\delta$. The quantity being represented is
\begin{equation}
|\mathcal{F}(\delta)|=\dfrac{1}{2}| \Tr \left[J_R^{-1}\left(\alpha \right) J(\delta)\right]| \ ,
\label{eq:fidelity+delta}
\end{equation}%
with
\begin{equation}
J(\delta)\equiv J_{\theta_3}\left( \pi(1+\delta) \right) J_{\theta_2}\left( -2\pi(1+\delta) \right) J_{\theta_1}\left( \pi(1+\delta) \right).
\label{eq:J_of_delta}
\end{equation}%
Note that here a unique value of the parameter $\delta$ can be considered for the three wave-plates provided that their dispersion is the same, which is the case for the system used in our experimental study.
Note also that it is sufficient to consider target rotation angles $|\alpha| \leq 90$ degrees because larger angles are redundant, also the situation for negative angles $\alpha$ is symmetric to the one for positive ones. First we show in Fig.~\ref{fig:fidelity}(a) the standard case where two HWPs under a relative angle $\alpha/2$ are used to create a rotator by an angle $\alpha$. While such a configuration acts as a perfect rotator at the central wavelength ($\mathcal{F}=1$ for $\delta=0$), for all four considered angles $\alpha$ between 15 and 90 degrees the fidelity is found to drop quite quickly as $\delta$ departs from zero. This means that such a structure is not spectrally robust. Interestingly, the robustness improves already significantly by means of a small modification, still using only two wave-plates. If the second HWP is turned by an additional 90 degrees, its retardation becomes negative ($=-\pi$) and, as seen in Fig.~\ref{fig:fidelity}(b), the function $\mathcal{F}(\delta)$ remains large over a much wider range of the parameter $\delta$. Figure~\ref{fig:fidelity}(c) show the case where the sequence of Fig.~\ref{fig:setup-1}(a) is implemented with $\varphi_1=\pi$, $\varphi_2=+ 2\pi$ and $\varphi_3=\pi$. This corresponds to the situation where the fast axes of the two HWP and of the central FWP are all oriented within an angle $\alpha/2$. In this case the retardation dispersion associated to the FWP reinforces the dispersion of the HWPs and the fidelity drops even faster than in the case of Fig.~\ref{fig:fidelity}(a). Finally, Fig.~\ref{fig:fidelity}(d) show our chosen configuration for which the orientation angles follow Eq.~(\ref{eq: angles relations}) meaning that the slow axis of the FWP is aligned in between the fast axes of the external HWPs. It is evident that in this case the function $\mathcal{F}(\delta)$ gets flatter on the top and is wider than in any other case in Fig.~\ref{fig:fidelity}. As we will discuss later, the experimentally most relevant range for the parameter $\delta$ is roughly $-0.5 \leq\delta \leq 0.5$, for which this three wave-plate configuration of Fig.~\ref{fig:fidelity}(d) is clearly outperforming any other case in Fig.~\ref{fig:fidelity}. It is also worth noting that the curve $\mathcal{F}(\delta)$ is found to be broader for small target rotation angles $\alpha$ than for larger ones. This situation is distinct from the one treated earlier by Rangelov et al. \cite{Rangelov15} using composite structures with a larger number of HWP ($\geq 6$), where the fidelity curve was found to be slightly wider for the the largest target rotation angles.

Even though the fidelity is high, when the wavelength departs from the central design wavelength $\lambda_0$, the wave-plate sequence will not act exactly as the nominally designed rotator. However, the function remains in very good approximation the one of a rotator. On the one hand it is impossible to avoid that the output wavelength would get a (small) elliptical polarization component for $\lambda \neq \lambda_0$. On the other hand the polarization rotation angle, characterized by the angle leading to maximum transmission through a properly oriented analyzer, will be a smooth function of $\delta$ (or $\lambda$). We discuss these issues for the cases in the right column of Fig.~\ref{fig:fidelity} for which the fidelity curves are broader. It is worth noticing that the configurations of the left column of Fig.~\ref{fig:fidelity} perform much worst also in this matter.

Figure~\ref{fig:rot-angle_theory} shows the theoretical expected rotation angle as a function of $\delta$ for the sequences used for Fig.~\ref{fig:fidelity}(b) and Fig.~\ref{fig:fidelity}(d). The depicted rotation angle is obtained by applying the matrix $J(\delta)$ in (\ref{eq:J_of_delta}) to an initial horizontal linear polarization and projecting the obtained Jones vector into an analyzer oriented at a variable angle $\gamma$. The angle $\gamma$ for which the transmission would be maximum (maximum squared projection) corresponds to the expected rotation angle. It is seen that in both cases the rotation angle corresponds to the target one at $\delta=0$ and decreases symmetrically for positive and negative values of $\delta$. In the most important case of three wave-plates the behavior is very well approximated (but not exactly given) by the function $\alpha(\delta)=\alpha(\delta=0)\cos^2(\delta \pi/2)$.
\begin{figure}[t!]
\includegraphics[width=\columnwidth]{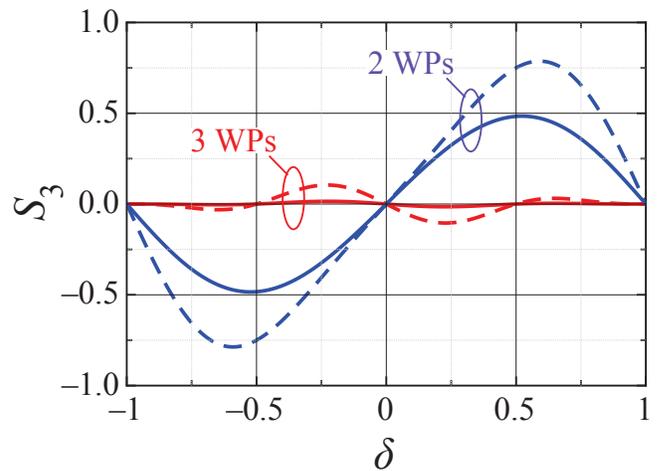}
    \caption{$S_3$-component of the output polarization Stokes vector for the two cases shown in Fig.~\ref{fig:rot-angle_theory}. The blue curves are for the two HWPs case of Fig.~\ref{fig:rot-angle_theory}(a), the red curves are for the three WP sequence of Fig.~\ref{fig:rot-angle_theory}(b). Solid curves are for a target rotation angle $\alpha=30$ deg, while dashed curves are for  $\alpha=60$ deg.}
  \label{fig:ellipticity}
\end{figure}

For the two HWP configuration the decrease of the rotation angle with increasing $|\delta|$ is initially less steep. However, for this configuration the output light field polarization exhibits a rather strong ellipticity. In contrast, for the three wave-plates sequence the output light remains always very close to a linear polarization state, what represent the major advantage of this configuration. To prove this we represent for both cases in Fig.~\ref{fig:ellipticity} the component $S_3$ of the Stokes vector \cite{Azzam} as a function of $\delta$. The $S_3$ component is defined as
\begin{equation}
S_3 \equiv I_R-I_L \ ,
\label{eq:S3}
\end{equation}%
where $I_R$ and $I_L$ are the normalized intensity transmissions of the output light field through a right-circular and a left-circular analyzer, respectively. Right and left circular polarized light have $S_3=+1$ and $S_3=-1$, respectively, while linear polarized light has $S_3=0$ and lies on the equator of the polarization Poincar\'{e} sphere. A scrutiny of Fig.~\ref{fig:ellipticity} clearly shows that the $S_3$ component remains always very small in the case of the three WP sequence treated in the present work, while significant values of $S_3$ are found for the sequence composed of two HWPs. By defining the degree of linear polarization as
\begin{equation}
\xi \equiv \sqrt{1-(S_3)^2}
\label{eq:degree-linear-pol}
\end{equation}%
we find that $\xi$ is always very close to one for the three WP sequence ($\xi > 99.9\%$ for $\alpha = 30$ deg, $\xi > 99.4\%$ for $\alpha = 60$ deg, and still $\xi > 95\%$ for the worst case of $\alpha = 90$ deg). In contrast, for the two HWP case, since the circular polarized component is much stronger, the degree of linear polarization can decrease to $\xi \approx 87.5\%$ for $\alpha = 30$ deg, to $\xi \approx 61.7\%$ for $\alpha = 60$ deg and to $\xi \approx 40\%$ for $\alpha = 90$ deg.

 It is worth to note that the curves for the rotation angle in Fig.~\ref{fig:rot-angle_theory} give the orientation of the longest main axis of the output polarization ellipse. Only in the case where the output light is still linear polarized, which is essentially true for Fig.~\ref{fig:rot-angle_theory}(b) as discussed above, this angle corresponds to the rotation angle of the linear polarization. In the specific case of Fig.~\ref{fig:rot-angle_theory}(a) the apparently broader curves with respect to Fig.~\ref{fig:rot-angle_theory}(b) are related to the strong ellipticity of the output wave for values of $\delta$ close to $\pm 0.5$ seen in Fig.~\ref{fig:ellipticity}. This pushes up artificially the apparent rotation angle around these values. This kind of artifact does not mean that the two WPs configuration would be better than the three WPs configuration. In fact, the fidelity, which is broader in the three WPs configuration, gives a much better measure of the usefulness of the device, as discussed later.

The above discussion clearly shows that, unlike the other configurations, the three wave-plate sequence with nominal retardations $(\varphi_1, \varphi_2, \varphi_3)=(\pi, -2\pi, \pi)$ offers a combination of a robust behavior under variation of $\delta$ and of the wavelength, as well as of an always nearly linearly polarized output light field. In the next section we will investigate this configuration experimentally.

\section{Experiments}
The experimental set-up for the characterization of our optical rotator is shown in Fig.~\ref{fig:setup-1}(b). The heart of the system is composed of two crystal polarizers (acting as polarizer and analyzer) surrounding the three wave-plate rotator sequence. The light source is either a monochromatic laser source (He-Ne laser at $\lambda=632.8$ nm) or a broadband white light source (Thorlabs SLS201L/M). Three distinct operation modes are used. In the case of the monochromatic He-Ne laser source the spectral filter and the mirror behind the analyzer are removed and the detection of the light transmitted through the analyzer occurs by means of a Si-photodiode. The same detection mode is used in the case where the broadband light source is combined with a spectral filter at different wavelengths to leave a quasi-monochromatic 10~nm wide (FWHM) spectrum through the system. Finally, by inserting the switchable mirror before the detector, the use of the broadband source without the spectral filter allows to detect the whole transmitted spectrum through the analyzer with the help of a spectrometer connected to a computer (OceanOptics USB4000-VIS-NIR).

The three wave-plates composing the rotator sequence are realized using three tunable liquid crystal (LC) retarders (Thorlabs LCC1413-A), whose retardations are adjusted with an external applied voltage. The voltage-retardation curve of each of the LC retarder has been priorly calibrated over the spectral range of interest using a Soleil-Babinet optical compensator (Thorlabs SBC-VIS) put in series with the LC retarder and whose mechanically adjustable retardation is known.

\begin{figure}[t!]
  \includegraphics[width=\columnwidth]{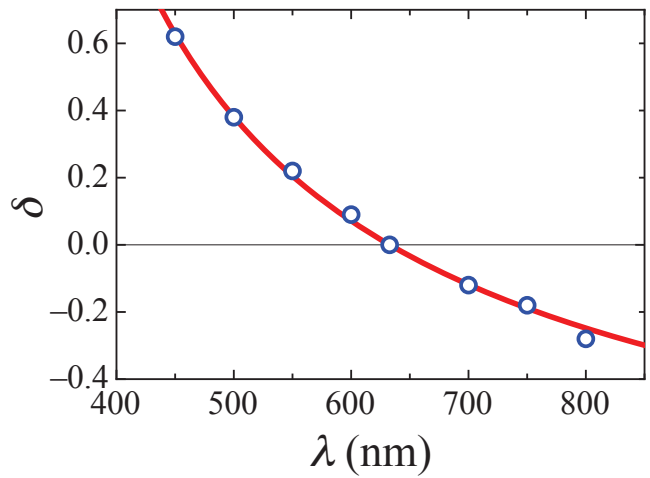}
  \caption{Experimental wave-plate relative retardation deviation $\delta$ as a function of $\lambda$ referred to the retardation at a central wavelength $\lambda_0=633$ nm (blue circles). The red fitted curve is according to (\ref{eq:delta_fit-of-dispersion}).}
  \label{fig: delta vs lambda}
\end{figure}

The above calibration allows to establish the LC WP retardation for each applied voltage at each of the filtered wavelengths ($450, 500, 550, 600, 633, 700, 750$ and $800$ nm) and to determine the spectral dependence of the relative deviation $\delta$ for a given central wavelength $\lambda_0$. The latter is determined by considering the wavelength dependence of the maximum retardation (maximum $\Delta n$) found for zero applied voltage. The obtained nonlinear relationship $\delta(\lambda)$ is shown in Fig.~\ref{fig: delta vs lambda} for the case where $\lambda_0= 633$ nm is the wavelength at which $\delta(\lambda_0)=0$. Clearly in this case the spectral range between 450 and 800 nm spans a $\delta$-range between $\approx 0.65$ and $-0.3$. The fitted dispersion curve in Fig.~\ref{fig: delta vs lambda} follows the relationship
\begin{equation}
\delta(\lambda)=\frac{\lambda_0^2-\bar{\lambda}^2}{\lambda^2-\bar{\lambda}^2}\frac{\lambda}{\lambda_0}-1  \  ,
\label{eq:delta_fit-of-dispersion}
\end{equation}%
with an effective oscillator wavelength $\bar{\lambda}=221.2$ nm. The above expression is obtained by assuming a simplified Sellmeier-like function for the dispersion of the LC birefringence $\Delta n=n_{\mathnormal{s}}-n_{\mathnormal{f}}$ entering Eq.~(\ref{eq:retardation}).

\begin{figure}[t!]
  \includegraphics[width=\columnwidth]{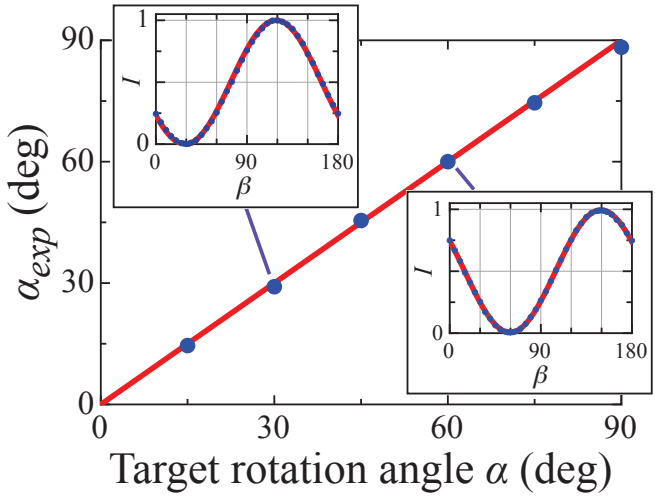}
  \caption{Experimentally measured rotation angle $\alpha_{exp}$ (blue dots) vs. the target rotation angle  $\alpha$ for the nominal wavelength $\lambda=\lambda_0=632.8$ nm. The two insets show the measured transmitted intensity $I$ through the analyzer oriented at an angle $\beta$ for $\alpha=30^\circ$ and $60^\circ$, as well as the function $I(\beta)$ given in the text. The orientation angles of the WPs are arranged according to Eq.~(\ref{eq: angles relations}).}
  \label{fig:experimental angles}
\end{figure}

We first analyze the behavior of the three-WP rotator at the wavelength $\lambda_0$ for which it is designed. In this case the He-Ne laser source at $\lambda_0=632.8$ nm is used and the LC wave-plates  are adjusted to be half- or full wave-plates at this wavelength. Figure~\ref{fig:experimental angles} compares the target rotation angles $\alpha$ with the experimentally measured polarization rotation angles and confirms an excellent agreement. The target rotation angles are adjusted by changing the orientations $\theta_1$ and $\theta_3$ according to (\ref{eq: angles relations}) and the experimental rotation angles $\alpha_{exp}$ are obtained by finding the minimum and maximum transmitted intensity upon rotation of the analyzer behind the rotator sequence in Fig.~\ref{fig:setup-1}(b). The insets in Fig.~\ref{fig:experimental angles} show examples of such measurements for $\alpha=30^\circ$ and $\alpha=60^\circ$. The dependence of the transmitted intensity $I(\beta)$ on the analyzer orientation $\beta$ (with $\beta=0$ being the extinction position in absence of the rotator sequence) is $I(\beta)=I_{0}\ \sin^2{(\beta-\alpha_{exp})}+I_{min}$, where $I_0$ is the modulation amplitude and $I_{min}$ is the minimum transmission. The full contrast seen in the insets in Fig.~\ref{fig:experimental angles} confirms that the output light maintains its linear polarization. Indeed, the degree of linear polarization $\xi$ in (\ref{eq:degree-linear-pol}) corresponds to the fringe visibility of such measurements given by $(I_{max}-I_{min})/(I_{max}+I_{min})=I_0/(I_0+2 I_{min})$.

\begin{figure}[t!]
   \includegraphics[width=\columnwidth]{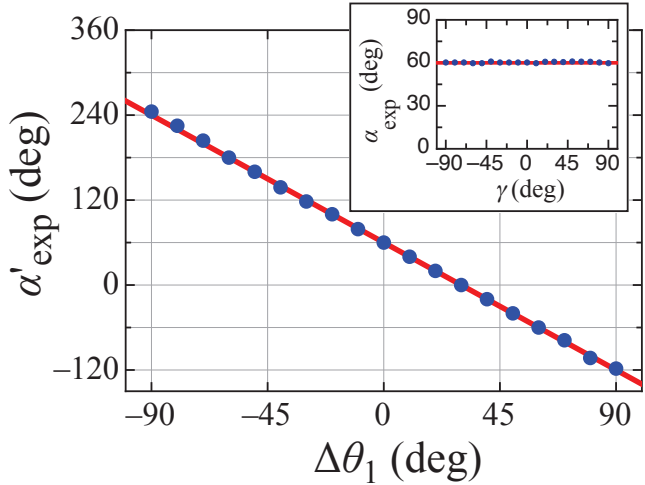}
  \caption{Tuning of the experimental rotation angle  $\alpha'_{exp}$ upon a variation $\Delta\theta_1$ of the orientation of the first HWP in the sequence and $\alpha=60^\circ$ for $\Delta\theta_1=0$. The solid red curve corresponds to the relation $\alpha'=\alpha-2\Delta\theta_1$. The inset shows the experimental rotation angle $\alpha_{exp}$ as a function of the input polarization angle $\gamma$ for the target rotation of $\alpha=60^\circ$. $\gamma=0$ corresponds to horizontal input polarization (the bisectrix of the HWP orientation angles), $\lambda_0=\lambda= 632.8$ nm. }
  \label{fig: wp1 rotation + initial polarization rotation}
\end{figure}

The rotation angle of the above WP sequence can be used as a tunable rotator by reorienting only one of the three wave-plates, the first or the third. Let us rotate the first wave plate, initially oriented under the angle $\theta_1$ by a supplementary angle $\Delta\theta_1$, so that $\theta'_1=\theta_1+\Delta\theta_1$. With $\alpha=2(\theta_3 - \theta_1)$ and $\alpha'=2(\theta_3 - \theta'_1)$ the new target rotation angle $\alpha'$ shall vary double as fast as $\theta'_1$, i.e. $\alpha'=\alpha-2\Delta\theta_1$.
This is confirmed in Fig.~\ref{fig: wp1 rotation + initial polarization rotation} that shows the tuned rotation angle $\alpha'$ for a variation of $\Delta\theta_1$ between $-90^\circ$ and $+90^\circ$ and an initial target angle $\alpha=60^\circ$.

Unlike the rotation of polarization by a single half-wave plate, which depends on the input polarization direction $\gamma$, a rotator should rotate the linear polarization by the same amount independently of $\gamma$. The inset in Fig.~\ref{fig: wp1 rotation + initial polarization rotation} shows the constancy of $\alpha_{exp}$ upon variation of $\gamma$ for the same rotator sequence used for Fig.~\ref{fig:experimental angles} and $\alpha=60^\circ$. The expected independence on the input polarization angle is therefore well verified experimentally.

\begin{figure}[t]
  \includegraphics[width=\columnwidth]{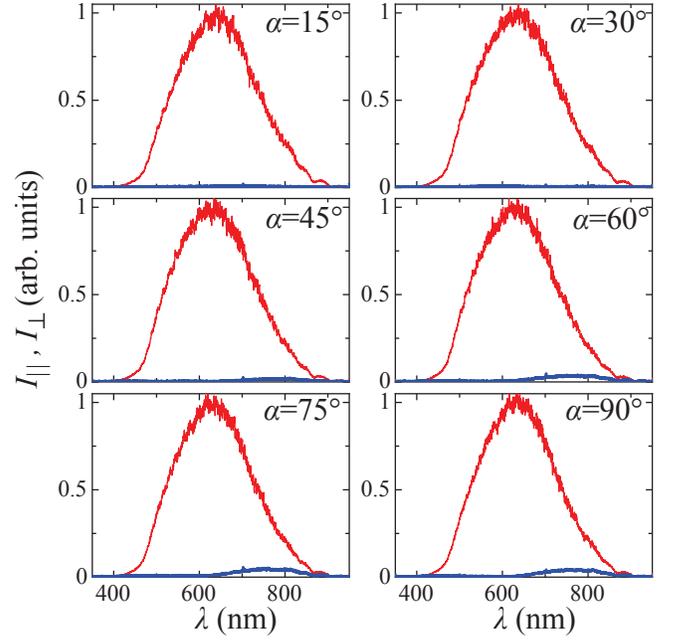}
  \caption{Spectra of the normalized intensities $I_{\parallel}$ (red curves) and $I_{\perp}$ (blue curves) transmitted through the analyzer in transmission, resp. extinction mode. For $I_{\parallel}$ the transmission direction of the analyzer corresponds to the target output polarization at the nominal wavelength $\lambda_0=550$ nm of the composite rotator.}
  \label{fig:spectroscopy}
\end{figure}

Next we test the broadband behavior and the robustness of the rotator if the used wavelength differs from the nominal wavelength $\lambda_0$. In this case, in order to better center the available spectrum into the corresponding range of the $\delta$-parameter, we choose  $\lambda_0=550$ nm.
First we send the whole spectrum of the broadband source through the composite rotator and detect the corresponding spectrum after passing the analyzer with the optical spectrometer (see Fig.~\ref{fig:setup-1}(b)).
The analyzer is put either in transmission mode (transmitted intensity = $I_{\parallel}$) or in extinction mode (transmitted intensity = $I_{\perp}$). In transmission mode the analyzer transmission axis is put parallel to the expected output polarization direction under the target rotation angle $\alpha$ for the nominal wavelength, while in extinction mode it is put perpendicular to this direction. Figure~\ref{fig:spectroscopy} shows the intensities $I_{\parallel}$ (red) and $I_{\perp}$ (blue) for six values of $\alpha$. While for small $\alpha$ the extinction in extinction mode is nearly perfect over the whole spectrum, for larger rotation angles a weak transmitted intensity persists at the border of the spectrum. This is connected to the narrower fidelity function $|\mathcal{F}(\delta)|$ for large than for small angles $\alpha$  seen in Fig.~\ref{fig:fidelity}(d).
\begin{figure}[t!]
  \includegraphics[width=\columnwidth]{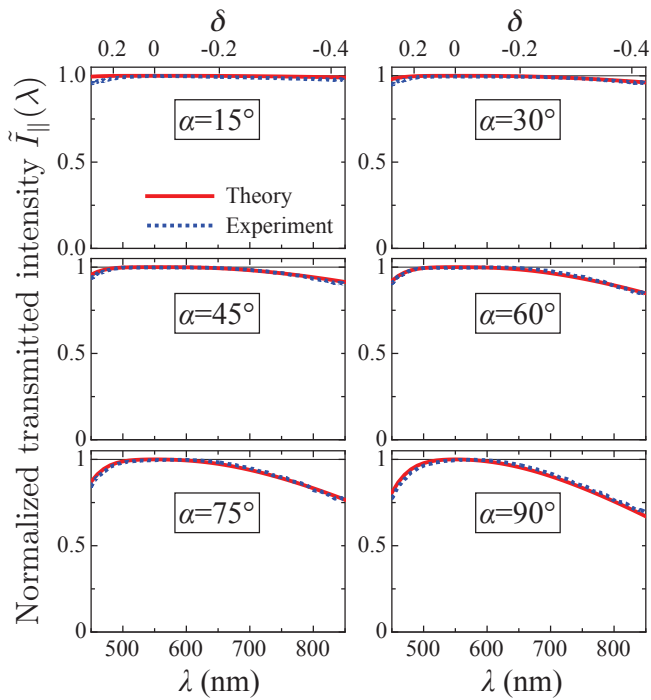}
    \caption{Normalized transmitted intensity spectra $\tilde{I}_{\parallel}(\lambda)$, the dotted blue curves give the experimental spectra while the underlying solid red lines are the theoretically expected spectra. The nominal wavelength for the three WPs rotator is $\lambda_0=550$ nm. The upper scale give the corresponding values for $\delta$.}
    \label{fig:spectroscopy normalized}
\end{figure}

The red curves for $I_{\parallel}$ in Fig.~\ref{fig:spectroscopy} do not permit a direct comparison with the theoretically expected intensity in transmission mode because they are influenced by the source spectrum and detector sensitivity, as well as the transmission spectrum of all optical elements in the set-up. To take away all these effects and permit this comparison we normalize therefore the intensity $I_{\parallel}$ as $\tilde{I}_{\parallel} \equiv I_{\parallel}/(I_{\parallel}+I_{\perp})$. Figure~\ref{fig:spectroscopy normalized} shows the corresponding spectra for $\tilde{I}_{\parallel}(\lambda)$ together with the theoretically expected ones. The latter are obtained by applying the resulting Jones matrix (\ref{eq:fidelity+delta}) to the input polarization Jones vector and projecting the resulting expected output polarization onto the analyzer to obtain the expected transmitted intensity as the square of the module of the projected polarization vector. The conversion between the wavelength $\lambda$ and the relative retardation deviation $\delta$ is made by means of Eq.~(\ref{eq:delta_fit-of-dispersion}) as discussed in connection to Fig.~\ref{fig: delta vs lambda}. As can be seen in Fig.~\ref{fig:spectroscopy normalized}, the agreement between the measured and the expected normalized transmission spectra is excellent. Even in the worst case scenario ($\alpha=90^\circ$ and a wavelength exceeding the nominal wavelength by 300 nm) the normalized transmitted intensity $\tilde{I}_{\parallel}$ is still $\approx 70\%$. A careful analysis shows that, in the case where the output light is still linearly polarized, the normalized transmitted intensity $\tilde{I}_{\parallel}$ corresponds to the square of the fidelity $|\mathcal{F}(\delta)|^2$,  what justifies the choice of the fidelity as the quantity to optimize.  In our case, by measuring the fringe visibility for wavelength filtered radiation as in the insets of Fig.~\ref{fig:experimental angles}, the degree of linear polarization $\xi$ is found to be close to 100$\%$ throughout, as expected theoretically (see Fig.~\ref{fig:ellipticity} and related discussion). Thus the above condition is satisfied and the measurements of Fig.~\ref{fig:spectroscopy normalized} can be considered as direct measurements of the fidelity squared.

\begin{figure}[t!]
  \includegraphics[width=\columnwidth]{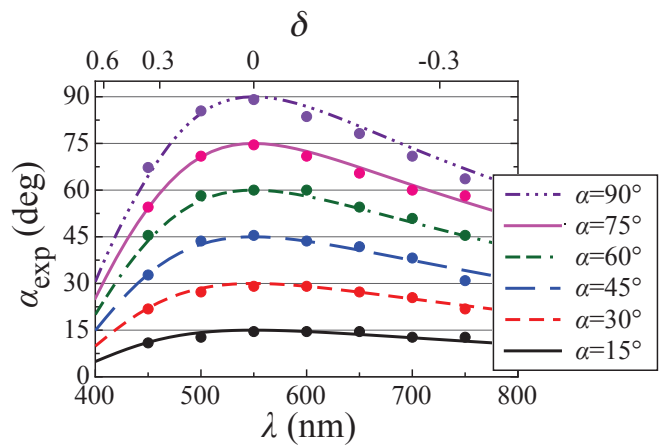}
    \caption{Variation of the polarization rotation angle $\alpha_{exp}$ with wavelength. The points are measured values at seven distinct wavelengths and the curves give the theoretically expected dependence. The $\alpha$-values characterizing each curve are the target rotation angles for the nominal central wavelength $\lambda_0=550$ nm. The upper abscissa give the values for $\delta$ corresponding to the wavelengths. }
    \label{fig:exp+theo rotation angles vs lambda}
\end{figure}

Therefore the decrease of $\tilde{I}_{\parallel}$ far from the nominal wavelength is not due to a decrease of $\xi$ but it is due to the dependence of the rotation angle on the parameter $\delta$ discussed in Fig.~\ref{fig:rot-angle_theory}(b). To check this we have performed measurements of the experimental rotation angle at seven filtered wavelengths ($\lambda=450, 500, 550, 600, 650, 700$ and $750$ nm) by inserting the corresponding spectral filter after the broadband light source in the set-up. The experimentally measured and theoretically expected rotation angles (see also Fig.~\ref{fig:rot-angle_theory}(b)) are shown in Fig.~\ref{fig:exp+theo rotation angles vs lambda} and show a very good agreement.

\section{Discussion and conclusion}
We have proposed theoretically and verified experimentally a new design for a composite broadband polarization rotator composed of only three wave-plates, two half-wave plates and one full-wave plate for the central nominal wavelength. This design is simpler with respect of earlier broadband composite rotators composed of a larger number of wave-plates \cite{Rangelov15, Dimova15, Stojanova2019}. We have shown that the output polarization state remains nearly linear even for strong departure from the nominal central wavelength $\lambda_0$ and that the polarization rotation angle has its maximum at $\lambda_0$ and diminishes smoothly away from this wavelength. The rotation angle can be tuned by rotating only one of the wave-plates and is robust against the initial polarization direction.  It is useful to compare the values of the figure-of-merit $Q$ defined in Eq.~(\ref{figure_of_merit}) for our selected configuration, the three other configurations of Fig.~\ref{fig:fidelity}, and the case of Ref.~\cite{Stojanova2019}, which gave the wider fidelity curves to date. By taking the example of a target rotation angle $\alpha=60^\circ$, the standard rotator composed of two HWPs of Fig.~\ref{fig:fidelity}(a) has a figure-of-merit $Q=0.504$, the addition of a FWP with the "wrong" sign of the phase shift as in Fig.~\ref{fig:fidelity}(c) leads to a similar value of $Q=0.491$. In contrast, the zero-pulse-area-like two-HWPs configuration of Fig.~\ref{fig:fidelity}(b), that we have discussed in larger extent, leads to an improvement since $Q=0.750$. However, as discussed above, this configuration is related to strongly elliptical polarized output light for a certain range of the detuning parameter $\delta$. Finally, the three-WPs design studied in this work has  $Q=0.808$. This value compares favorably with the corresponding values for the schemes of Ref.~\cite{Stojanova2019}, which are 0.812; 0.720; 0.863 and 0.727 for a total number of WPs $N$ equal to 4; 6; 8 and 10, respectively. Only the case $N=8$ is associated to a significantly better figure-of-merit than for the present much simpler configuration.

In conclusion, we believe that the three wave-plates rotator discussed in this work will  be useful for light polarization management of broadband, short-pulse or tunable light sources.

\acknowledgements
This work is supported by EU Horizon-2020 ITN project LIMQUET (contract number 765075) and by the Bulgarian Science Fund Grant No. DN 18/14.


\begin{thebibliography}{99}

\bibitem{Azzam} M. A. Azzam, N. M. Bashara, Ellipsometry and Polarized
Light, North Holland, Amsterdam, 1977.

\bibitem{Goldstein} D. H. Goldstein, Polarized Light, 3rd Ed., (CRC Press, Boca Raton, 2011).

\bibitem{Duarte} F. J. Duarte, Tunable Laser Optics, 2nd Ed., (CRC Press, Boca Raton, 2015).

\bibitem{Kliger-book} D. S. Kliger, J. W. Lewis, C. E. Randall, Polarized Light in Optics and Spectroscopy,(Academic Press, Boston, 1990).

\bibitem{West49} C. D. West, and A. S. Makas, The spectral dispersion of birefringence, especially of birefringent plastic sheets, J. Opt. Soc. Am. \textbf{39}, 791-794 (1949).

\bibitem{Destriau49} M. G. Destriau, and J. Prouteau, R{\'e}alisation d'un quart d'onde quasi achromatique par juxtaposition de deux lames cristallines de m{\^e}me nature, J. Phys. Radium \textbf{10}, 53-55 (1949).

\bibitem{Pancharatnam55-1} S. Pancharatnam, Achromatic combinations of birefringent plates. Part I. An achromatic circular polarizer, Proc. Indian Acad. Sci. A \textbf{41}, 130-136 (1955).

\bibitem{Pancharatnam55-2} S. Pancharatnam, Achromatic combinations of birefringent plates. Part II. An achromatic quarter-wave plate, Proc. Indian Acad. Sci. A \textbf{41}, 137-144 (1955).

\bibitem{McIntyre68} C. M. McIntyre, and S. E. Harris, Achromatic wave plates for the visible spectrum, J. Opt. Soc. Am. \textbf{58}, 1575-1580 (1968).

\bibitem{Ardavan07} A. Ardavan, Exploiting the Poincar{\'e}--Bloch symmetry to design high-fidelity broadband composite linear retarders, New J. Phys. \textbf{9},  24 (2007).

\bibitem{Ivanov2012} S. S. Ivanov, A. A. Rangelov, N. V. Vitanov, T. Peters, and T. Halfmann, Highly efficient broadband conversion of light polarization by composite retarders J. Opt. Soc. Am. A \textbf{29}, 265-269 (2012).

\bibitem{Peters2012} T. Peters, S. S. Ivanov, D. English, A. A. Rangelov, N. V. Vitanov and T. Halfmann, Variable ultrabroadband and narrowband composite polarization retarders, Appl. Opt. \textbf{51}, 7466-7474 (2012).

\bibitem{Dimova2014} E. Dimova, S. S. Ivanov, G. Popkirov, and N. V. Vitanov, Highly efficient broadband polarization retarders and tunable polarization filters made of composite stacks of ordinary wave plates, J. Opt. Soc. Am. A \textbf{31}, 952-956 (2014).

\bibitem{Dimova2016} E. Dimova, W. Huang, G. Popkirov, A. A. Rangelov, and E. Kyoseva, Broadband and ultra-broadband modular half-wave plates, Opt. Commun. \textbf{366}, 382-385 (2016).

\bibitem{Levitt86} M. H. Levitt, Composite pulses, Progr. Nucl. Magn. Reson. Spectrosc. \textbf{18}, 61-122 (1986).

\bibitem{Rangelov15} A. A. Rangelov and E. Kyoseva, Broadband composite polarization rotator, Opt. Commun. \textbf{338}, 574-577
(2015).

\bibitem{Dimova15} E. Dimova, A. A. Rangelov, and E. Kyoseva, Tunable bandwidth optical rotator, Photon. Res. \textbf{3}, 177-179
(2015).

\bibitem{Stojanova2019} E. Stojanova, M. Al-Mahmoud, H. Hristova, A. A. Rangelov, E. Dimova, and N. V. Vitanov, Achromatic polarization rotator with tunable rotation angle, J. Opt. \textbf{21}, 105403
(2019).

\bibitem{Moeller-book} K. D. M\"oller, Optics, 3rd Ed., (Univ. Science Books, Mill Valley, 1988).

\bibitem{Zhuang2000} Z. Zhuang, Y. J. Kim, and J. S. Patel, Achromatic linear polarization rotator using twisted nematic liquid crystals, Appl. Phys. Lett. \textbf{76}, 3995-3997
(2000).

\bibitem{Chung18} T. Y. Chung, M. C. Tsai, C. K. Liu, J. H. Li, and K. T. Cheng, Achromatic linear polarization rotators by tandem twisted nematic liquid crystal cells, Sci. Rep. \textbf{8}, 13691
(2018).

\bibitem{Messaadi} A. Messaadi, M. M. Sanchez-Lopez, A. Vargas, P.
Garcia-Martinez, and I. Moreno, Achromatic linear retarder with tunable retardance, Opt. Lett. \textbf{43}, 3277-3280 (2018).

\bibitem{Fabricius} H. Fabricius, Achromatic prism retarder for use in polarimetric sensors, Appl. Opt. \textbf{30}, 426-429 (1991).

\bibitem{Ye} C. Ye, Construction of an optical rotator using quarter-wave plates and an optical retarder, Opt. Eng. \textbf{34}, 3031-3035 (1995).

\bibitem{Davis} J. A. Davis, D. E. McNamara, D. M. Cottrell, and T.
Sonehara, Two-dimensional polarization encoding with a phase-only liquid-crystal spatial light modulator, Appl. Opt. \textbf{39}, 1549-1554 (2000).

\bibitem{Vasilev06} G. S. Vasilev, and N. V. Vitanov, Complete population transfer by a zero-area pulse,
Phys. Rev. A \textbf{73},  023416 (2006).

\bibitem{Torosov14} B. T. Torosov, and N. V. Vitanov, High-fidelity error-resilient composite phase gates,
Phys. Rev. A \textbf{90},  012341 (2014).



\end{thebibliography}
\end{document}